\documentclass[12pt]{article}
\usepackage{graphicx}
\usepackage{amssymb,amsmath}
\newcommand{\mvec}[1]{\boldsymbol{#1}}
\newcommand{\bnabla}{\boldsymbol{\nabla}}
\newcommand{\umu}{\mu}
\newcommand{\upi}{\pi}
\newcommand{\uepsilon}{\epsilon}
\newcommand{\mten}[1]{\mathbf{#1}}
\newcommand{\bes}{\begin{eqnarray}}
\newcommand{\ees}{\end{eqnarray}}
\author{Robert W. Johnson \\
\small Alphawave Research\\[-0.8ex]
\small Atlanta, GA, USA\\
\small \texttt{robjohnson@alphawaveresearch.com}\\}
\title{Macroscopic electromagnetic stress tensor for ionized media}
\date{\today\\
\small PACS: 52.25.Xz, 52.55.Dy, 47.65.Cb}

\newcommand{\citep}[1]{\cite{#1}}
\newcommand{\beq}{\begin{equation}}
\newcommand{\eeq}{\end{equation}}
\newcommand{\bea}{\begin{eqnarray}}
\newcommand{\eea}{\end{eqnarray}}
\newcommand{\mcal}[1]{\mathcal{#1}}
\newcommand{\mbb}[1]{\mathbb{#1}}
\newcommand{\mbf}[1]{\mathbf{#1}}
\newcommand{\mrm}[1]{\mathrm{#1}}
\newcommand{\del}{\bnabla}
\newcommand{\divr}{\bnabla \cdot}
\newcommand{\curl}{\bnabla \times}
\newcommand{\epsi}{\uepsilon}

\newcommand{\phihat}{\boldsymbol{\hat{\phi}}}
\newcommand{\Zhat}{\boldsymbol{\hat{Z}}}
\newcommand{\Rhat}{\boldsymbol{\hat{R}}}

\newcommand{\half}{\frac{1}{2}}
\newcommand{\wt}[1]{\widetilde{#1}}

\newcommand{\oover}[1]{\dfrac{1}{#1}}

\newcommand{\ptild}{\widetilde{p}}

\newcommand{\parxy}[2]{\dfrac{\partial\, {#1}}{\partial {#2}}}

\newcommand{\psub}[1]{\partial_{#1}}
\newcommand{\Tr}{\mathrm{Tr}}

\begin{document}

\maketitle

\begin{abstract}
Following the arguments presented by Mansuripur [Opt. Express {\bf 16}, 14821-14835 (2008)], we suggest a form for the macroscopic electromagnetic stress tensor appropriate for ionized media.  The generalized Lorentz force includes the effects of polarization forces as well as those on the free charge and current densities.  The resulting tensor is written in terms of the fields $\mvec{D}$, $\mvec{B}$, $\mvec{E}$, and $\mvec{H}$.  Its expression for a fully ionized medium subject to an external electromagnetic field is discussed, as are the plasma conservation equations.  An apparatus is suggested for its experimental discrimination.
\end{abstract}


\section{Introduction}\label{sec:intro}
While the microscopic form of the electromagnetic stress tensor is well known \citep{jackson-third}, its macroscopic form continues to be a matter of debate \citep{loudon-063802,liu-10261L,engel-428E}.  Of interest are the effects of the material polarization fields $\mvec{P}$ and $\mvec{M}$, particularly the magnetic contribution which has variously been given \citep{melcher-81,rosen-82} by Lorentz and Kelvin as $\mbf{F}_\mathrm{LK} = \mu_0 \mbf{J} \times \mbf{H} + \mu_0 \mbf{M} \cdot \del \mbf{H}$ and by Korteweg and Helmholtz as $\mbf{F}_\mathrm{KH} = \mbf{J} \times \mbf{B} - \mbf{H} \cdot \mbf{H} \del \mu / 2$.  Following the arguments presented by Mansuripur \citep{mansuripur-1608,mansuripur-1619}, we propose a form for the macroscopic electromagnetic stress tensor for ionized media expressed in terms of the fields $\mvec{D}$, $\mvec{B}$, $\mvec{E}$, and $\mvec{H}$.  As no assumption on the form of the constitutive relations is taken other their definition, we believe this expression has applicability beyond that of linear, homogeneous media.

With restriction to a fully ionized medium $\mvec{P} \rightarrow 0$, we evaluate the nonlinear relation between $\mvec{M}$ and $\mvec{H}$ explicitly.  For vanishing charge density,  the stress tensor may be expressed in terms of the pressure $p$, the magnetic field $\mvec{H}$, and the applied field $\mvec{E}$.  The electromagnetic stress may then be combined with the hydrodynamic stress to yield the equation for the net conservation of momentum.  With reduction to a stationary plasma carrying no free current in the absence of gravity, there remains a balance of pressure against the magnetic polarization force, suggesting a means for an experimental apparatus to determine whether its effect is present in a fully ionized medium.

\section{Field equations}\label{sec:field}
The macroscopic inhomogeneous Maxwell equations for the continuum (fluid) description of electromagnetic phenomena are \beq
\divr \mvec{D} = \rho_\mrm{f} \;, \;\;\; \curl \mvec{H} - \psub{t} \mvec{D} = \mvec{J}_\mrm{f} \;, 
\eeq where $\psub{t} \equiv \partial / \partial t$ and the sources appear opposite the fields.  The corresponding homogeneous equations are \beq
\divr \mvec{B} = 0 \;, \;\;\; \curl \mvec{E} + \psub{t} \mvec{B} = 0 \;,
\eeq and we emphasize that the proper interpretation of these two equations \citep{ryder-qft,davis70} is {\it not} as dynamical equations but rather as geometric constraints \citep{naka-798212,ward-1286287} yielding the potential formulation $\mvec{E} = - \del \varPhi - \psub{t} \mvec{A}$ and $\mvec{B} = \curl \mvec{A}$, and the gauge condition $\divr \mvec{A} + \umu_0 \epsi_0 \psub{t} \varPhi = 0$ expresses the continuity of the potential \citep{rouss-49249} associated with the continuity of the source $\divr \mvec{J} + \psub{t} \rho = 0$.  The constitutive relations are taken in their most general form, \beq
\mvec{D} - \mvec{P} = \epsi_0 \mvec{E} \;, \;\;\; \mvec{B} - \mvec{M} = \umu_0 \mvec{H} \;,
\eeq without assumption on their detailed expression.  Note the units $\mvec{M} = \umu_0 \mvec{M}_\mrm{SI}$ used in this section and the next.

In the following, the operator $\del$ is allowed to act upon the material polarizations $\mvec{P}$ and $\mvec{M}$.  Such allowance is consistent with the expressions $\curl \mvec{P} = \curl \mvec{D}$ and $\divr \mvec{M} / \umu_0 = - \divr \mvec{H}$ necessary to compute the extruded fields from the bar electret and bar magnet \citep{griffiths-89}.  No requirement of linearity is imposed, as the tensor structure developed accommodates naturally any variation in the material polarizations.

\section{Generalized Lorentz force}\label{sec:force}
In its microscopic form, the Lorentz force density is written as \beq
\mvec{F}_\mrm{L} = \sum_q \mvec{f}_q = \sum_q e_q (\mvec{E} + \mvec{v}_q \times \mvec{B}) \;,
\eeq where the sum is over the particle label $q$ and the fields $\mvec{E}$ and $\mvec{B}$ are to be evaluated at the particle position.  The most common macroscopic generalization is \beq
\mvec{F}_\mrm{L} = \sum_s n_s e_s (\mvec{E} + \mvec{V}_s \times \mvec{B}) \rightarrow \rho \mvec{E} + \mvec{J} \times \mvec{B} \;,
\eeq where the sum is over the species label $s$ and the fields $\mvec{E}$ and $\mvec{B}$ are averaged over the infinitesimal volume (fluid) element containing sufficient particle number for continuum thermodynamics to be applicable.  An averaging process is inherent in taking the particle velocity $\mvec{v}_q$ over to the continuum velocity $\mvec{V}_s = \sum_{q \in s} \mvec{v}_q / n_s$ which here includes any diamagnetic gyration.  The standard reduction gives the force density $\mvec{F}_\mrm{L} = \divr \mten{T}_\mrm{EB} - \epsi_0 \umu_0 \psub{t} \mvec{S}_{EB}$ in terms of the stress tensor $\mten{T}_\mrm{EB} = \epsi_0 (\mvec{E} \mvec{E} - \mten{I}\, E^2 / 2) + \umu_0^{-1} (\mvec{B} \mvec{B} - \mten{I}\, B^2 / 2)$ and the Poynting vector $\mvec{S}_\mrm{EB} = \mvec{E} \times \mvec{B} / \umu_0 \equiv \mrm{c}^2 \mvec{K}_{EB}$ for vacuum light speed $\mrm{c}$.  In polarizable media \citep{jackson-third,griffiths-89}, one uses the Poynting vector $\mvec{S}_\mrm{EH} = \mvec{E} \times \mvec{H}$ which accounts for the work done (energy deposited) to the free source density, suggesting that the most appropriate fields for the macroscopic description are $\mvec{E}$ and $\mvec{H}$.

We call the reader's attention to the excellent analysis by Mansuripur \citep{mansuripur-1608,mansuripur-1619}, particularly the statement that ``a new term had to be introduced to account for the force experienced by magnetic dipoles.''  This magnetic polarization force is familiar to anyone who has ever held two pieces of permanent magnet close together, accounting for the attraction or repulsion.  On energetic grounds \citep{mattuck-1976,aipdeskref}, one expects an isolated dipole $\mvec{m}$ to experience a force \beq \label{eqn:delmB}
\del (\mvec{m} \cdot \mvec{B}) = \mvec{m} \times (\curl \mvec{B}) + (\mvec{m} \cdot \del) \mvec{B} 
\eeq which accounts for the interaction with both currents and fields, and similarly for the electric polarization.  Mansuripur \citep{mansuripur-1608,mansuripur-1619} has recently presented an analysis of the generalized Lorentz force for polarizable media in the absence of sources, \beq
\mvec{F}_\mrm{media} = (\mvec{P} \cdot \del) \mvec{E} - (\psub{t} \mvec{M}) \times \epsi_0 \mvec{E} + (\mvec{M} \cdot \del) \mvec{H} + (\psub{t} \mvec{P}) \times \umu_0 \mvec{H} \;,
\eeq and much of the following as regards the media contribution appears there first.  To this expression we add the source contribution yielding the net Lorentz force $\mvec{F}_\mrm{L} = \mvec{F}_\mrm{media} + \mvec{F}_\mrm{free}$, where \beq
\mvec{F}_\mrm{free} = \rho_\mrm{f} \mvec{E} + \mvec{J}_\mrm{f} \times \umu_0 \mvec{H} \;.
\eeq  Why no $\mvec{J}_\mrm{f} \times \mvec{M}$ term?  From Eqn.~(\ref{eqn:delmB}) we see that such effect will be accounted for by the polarization term; furthermore, its additional presence would spoil the symmetry of the expressions below, hanging off the end like an unwanted tail.

One proceeds by expressing $\mvec{F}_\mrm{media}$ in terms of the material polarization tensors $\mvec{P} \mvec{E}$ and $\mvec{M} \mvec{H}$ using the identity $\divr (\mvec{P} \mvec{E}) = (\divr \mvec{P}) \mvec{E} + (\mvec{P} \cdot \del) \mvec{E}$ to write \bes \label{eqn:FPH}
\mvec{F}_\mrm{media} =& \divr (\mvec{P} \mvec{E}) - (\divr \mvec{P}) \mvec{E} - (\psub{t} \mvec{M}) \times \epsi_0 \mvec{E} \\
 & + \divr (\mvec{M} \mvec{H}) - (\divr \mvec{M}) \mvec{H} + (\psub{t} \mvec{P}) \times \umu_0 \mvec{H} \;,
\ees then substituting $\psub{t} \mvec{P} = \curl \mvec{H} - \mvec{J}_\mrm{f} - \epsi_0 \psub{t} \mvec{E}$ and $\psub{t} \mvec{M} = - \curl \mvec{E} - \umu_0 \psub{t} \mvec{H}$ along with $\divr \mvec{P} = \rho_\mrm{f} - \epsi_0 \divr \mvec{E}$ and $\divr \mvec{M} = - \umu_0 \divr \mvec{H}$ yields \bes
\mvec{F}_\mrm{media} =& \divr (\mvec{P} \mvec{E}) + \epsi_0 \left[ (\divr \mvec{E}) \mvec{E} + (\curl \mvec{E}) \times \mvec{E} \right] \\
 & + \divr (\mvec{M} \mvec{H}) + \umu_0 \left[ (\divr \mvec{H}) \mvec{H} + (\curl \mvec{H}) \times \mvec{H} \right] \\
 & - \epsi_0 \umu_0 \psub{t} (\mvec{E} \times \mvec{H}) - \rho_\mrm{f} \mvec{E} - \mvec{J}_\mrm{f} \times \umu_0 \mvec{H} \;.
\ees  Now adding the source term $\mvec{F}_\mrm{free}$ and expressing in dyads gives \bes
\mvec{F}_\mrm{L} =& \divr (\mvec{P} \mvec{E}) + \epsi_0 \left[ \divr (\mvec{E} \mvec{E}) - \del E^2 / 2 \right] \\
 & + \divr (\mvec{M} \mvec{H}) + \umu_0 \left[ \divr (\mvec{H} \mvec{H}) - \del H^2 / 2 \right] \\
 & - \epsi_0 \umu_0 \psub{t} (\mvec{E} \times \mvec{H}) \;,
\ees and one final substitution yields \bes
\mvec{F}_\mrm{L} &= &\divr (\mvec{D} \mvec{E}) + \divr (\mvec{B} \mvec{H}) - \del (\epsi_0 E^2 + \umu_0 H^2) / 2 - \epsi_0 \umu_0 \psub{t} \mvec{S}_\mrm{EH} \\
 &\equiv &\divr \mten{T}_\mrm{EH}^\mrm{DB} - \psub{t} \mvec{K}_\mrm{EH} 
\ees as the macroscopic electromagnetic stress tensor for source bearing polarizable media in terms of the fields $\mvec{D}$, $\mvec{B}$, $\mvec{E}$, and $\mvec{H}$, noting that $\epsi_0 E^2 = (\epsi_0 \mvec{E}) \cdot \mvec{E} = (\mvec{D} - \mvec{P}) \cdot \mvec{E}$ and similarly for $\umu_0 H^2$.  One may compare this expression with those by Abraham \citep{abraham-3033} and Minkowski \citep{minkowski-68472} as well as the method by Louden {\it et al} \citep{loudon-063802}.  This tensor form benefits from the verification of energy and momentum conservation by Mansuripur \citep{mansuripur-1608,mansuripur-1619} in a variety of situations.

Mansuripur considers also a media force which neglects the material polarization tensors $\mvec{P} \mvec{E}$ and $\mvec{M} \mvec{H}$ in Eqn.~(\ref{eqn:FPH}), \beq
\mvec{F}_\mrm{media}^\mrm{\,alt} = - (\divr \mvec{P}) \mvec{E} - (\psub{t} \mvec{M}) \times \epsi_0 \mvec{E}  - (\divr \mvec{M}) \mvec{H} + (\psub{t} \mvec{P}) \times \umu_0 \mvec{H} \;,
\eeq which Louden and Barnett \citep{louden-1411855L} show yields an equivalent force given certain conditions on the material and may be expressed as $\divr \mten{T}_\mrm{EH} - \psub{t} \mvec{K}_\mrm{EH}$ where $\mten{T}_\mrm{EH} = \mten{T}_\mrm{EH}^\mrm{DB} - \mvec{P} \mvec{E} - \mvec{M} \mvec{H}$.  Mansuripur \citep{mansuripur-1619} gives the corresponding torque density expressions \bes
\mvec{T}_\mrm{L} (\mvec{r}) &= &\mvec{r} \times \mvec{F}_\mrm{media} + \mvec{P} \times \mvec{E} + \mvec{M} \times \mvec{H} \;, \\
\mvec{T}_\mrm{L}^\mrm{alt} (\mvec{r}) &= &\mvec{r} \times \mvec{F}_\mrm{media}^\mrm{alt} \;,
\ees and we see the difference in the expressions for the torque about an observer $\mvec{r} \rightarrow 0$ is $\mvec{P} \times \mvec{E} + \mvec{M} \times \mvec{H}$, which may be related to the isolated dipole torques $\mvec{t}_p = \mvec{p} \times \mvec{E}$ and $\mvec{t}_m = \mvec{m} \times \mvec{B}$.  These torques serve to align the material polarizations along the fields in a fluid medium.

\section{Plasma magnetization}\label{sec:plasma}
Considering now one's favourite fully ionized plasma with species $s \in \{e,i\}$ and ion charge $e_i = z_i e$, we take $\mvec{P} \rightarrow 0$ such that $\mvec{D} = \epsi_0 \mvec{E}$, noting that the presence of neutrals and partials may contribute to an electric polarization $\mvec{P}$.  With a total particle number of $n \equiv n_e + n_i$, the electron density may be written $n_e = n_0 - \rho_\mrm{f} / e$, where $n_0 \equiv n_i z_i$ is the number of electrons required for charge neutrality.  Assuming equipartition of internal energy $W^\perp_s = 2 W^\parallel_s = T_s$ where $T_s \leftarrow k_B T_s$, the kinetic pressure is $p \equiv n\,T = \sum_s n_s T_s$ and the mass density $\rho_\mrm{m} \equiv n\,m = \sum_s n_s m_s$, exhibiting the utility of distinguishing the expressions $n$ and $n_0$.  For neutral, hydrogenic plasma, $z_i = 1$ and $n = 2 n_0$.   The free sources are defined by $\rho_\mrm{f} \equiv \sum_s n_s e_s \rightarrow 0$ and $\mvec{J}_\mrm{f} \equiv \sum_s n_s e_s \mvec{V}_s$, and the fluid momentum by $\mvec{K}_\mrm{f} \equiv \rho_\mrm{m} \mvec{V}_\mrm{f} = \sum_s n_s m_s \mvec{V}_s$.  From these relations, one may write $\mvec{V}_{i,e} = \mvec{V}_\mrm{f} \pm m_{e,i} \mvec{J}_\mrm{f} / e \rho_\mrm{m}$.  The velocity $\mvec{V}_s$ here is understood not to include the diamagnetic gyration $\mvec{J}_\mrm{dia} = \curl \mvec{M}_\mrm{SI}$ which is associated with the diamagnetic momentum $\mvec{K}_\mrm{dia}$.  In terms of the gyro-vector $\boldsymbol{\omega}_s \equiv - e_s \mvec{B}_s / m_s$, one writes $\mvec{K}_\mrm{dia} \equiv \rho_\mrm{m} \mvec{V}_\mrm{dia} = \curl \mvec{L}_\mrm{dia}$ for angular momentum $\mvec{L}_\mrm{dia} \equiv \sum_s n_s \langle \mvec{l}_s \rangle$, where $\langle \mvec{l}_s \rangle = 2 T_s \boldsymbol{\omega}_s / \omega_s^2$.

We now shift our notation so that the magnetic material polarization is in SI units $\mvec{M}_\mrm{SI} \leftarrow \mvec{M} / \umu_0$ and drop the subscript.  The species dipole moment per unit volume is taken as $\mvec{M}_s \equiv n_s \langle \mvec{m}_s \rangle = -n_s (W^\perp_s / B_s^2) \mvec{B}_s$, where the field felt by a single particle of species $s$ within the unit of volume is the net field less the particle's own contribution $\mvec{B}_s / \mu_0 \equiv \mvec{H} + \mvec{M}- \langle \mvec{m}_s \rangle = \mvec{H}+\mvec{M}_k + \alpha_s \mvec{M}_s$, where $k \neq s$ and $\alpha_s \equiv (n_s - 1)/n_s = 1-1/n_s$ is a unit-less factor.  For a sufficiently dense plasma, $\alpha_s \rightarrow 1$ and $\mvec{B}_s \rightarrow \mvec{B}$.  The material polarization is given by the net dipole density \citep{hazeltine-04}, which for $\ptild \equiv p / \mu_0$ and $\mvec{h} \equiv \mvec{H}/H$ may be written \bea 
\mvec{M} &\equiv& \sum_s \mvec{M}_s = n_e \langle \mvec{m}_e \rangle + n_i \langle \mvec{m}_i \rangle \;, \\
&=& - \sum_s \left( \ptild_s / \vert \mvec{H}+\mvec{M} \vert^2 \right) \left( \mvec{H}+\mvec{M} \right) \;, \\
&=& - \mvec{h} M  = - \mvec{h}\, \ptild / \left(H-M\right) \;,
\eea and has the physical solution $M/H = (1 - \sqrt{1 - 4 \ptild / H^2})/2$ as the plasma is diamagnetic \citep{marshall-122367}.  That simple expression represents a significant result, as we are aware of no other author investigating its utility.  Ultimately, the proper treatment of magnetization requires the use of quantum theory, in particular as to account for spin \citep{halzenmartin}. 

 From the form of the solution for $M$ one can immediately read a limit on the ratio of kinetic to free magnetic pressure, $\beta_H \equiv 2 \ptild / H^2 \leq 1/2$ for a dense plasma. In terms of the net field $B$, we have $\beta_B \equiv 2 \ptild / (H-M)^2 = \beta_H / (1 - M/H)^2 \leq 2$, and the ratio $M/H$ is limited to $1/2$.  In the dilute fluid limit $n_0 \rightarrow 1/r_i^3$ such that $\alpha \equiv (n-1)/n = 1 - 1/2 n_0 \rightarrow 1/2$, we find $\mvec{M}_s \rightarrow \mvec{m}_s$ so that the limits $\beta_H \rightarrow M/H \rightarrow 1$ when $T_i = T_e$.  One must be careful to define the appropriate unit of volume for a dilute plasma, which we feel should be on the order of the cube of the ion gyro-radius $r_i =  \omega_i^{-1} \sqrt{2 T_i / m_i}$.

The magnetized plasma stress tensor for a fully ionized medium, $\mten{T}^{EB}_{EH} = \mten{T}^{DB}_{EH}$ for $\mvec{P} \rightarrow 0$, may then be expressed as \beq
\mten{T}^{EB}_{EH} = \epsi_0 (\mvec{E} \mvec{E} - \mten{I} E^2 / 2) + \umu_0 [\mvec{H} \mvec{H} (1 + \sqrt{1 - 4 \ptild / H^2}) - \mten{I} H^2] / 2 \;,
\eeq where the electric field is the sum of that from plasma sources and any applied field $\mvec{E} = \mvec{E}_p + \mvec{E}_\mrm{app}$.  In the neutral limit $\divr \mvec{E}_p \rightarrow 0$, the plasma electric field is dynamic, $\curl \mvec{E}_p = - \umu_0 \psub{t} (1 - M/H) \mvec{H}$, thus specifying the tensor in terms of the pressure $p$ and the fields $\mvec{H}$ and $\mvec{E}_\mrm{app}$.

\section{Conservation equations}\label{sec:conserv}
Most analyses of plasma \citep{dendybook-93,staceybook05,fitzpatrick-notes} rely on the collisional Boltzmann equation \bes
\parxy{f}{t} + \mvec{v} \cdot \parxy{f}{\mvec{x}} + \mvec{a} \cdot \parxy{f}{\mvec{v}} &=& \left( \parxy{f}{t} \right)_\mrm{C} + \left( \parxy{f}{t} \right)_\mrm{S} \;, \\
 &\equiv& C + S \;,
\ees where $C$ and $S$ represent the collision and source terms whose moments $C_k$, $S_k$ are indexed by the natural numbers $k \in \mbb{N}_0$.  The species individually satisfy the particle continuity equation $\psub{t} n_s + \divr n_s \mvec{V}_s = S_{s0}$ where $\dot{n}_s \equiv [\mrm{d}_s + (\divr \mvec{V}_s)] n_s$ is the particle source rate and $\mrm{d}_s \equiv \psub{t} + \mvec{V}_s \cdot \del$ is the convective derivative, thus yielding the conservation of mass \beq
\psub{t} \rho_\mrm{m} + \divr \mvec{K}_\mrm{f} = \dot{\rho}_\mrm{m} \;,
\eeq where $\dot{\rho}_\mrm{m} \equiv \sum_s \dot{\rho}_s = \sum_s \dot{n}_s m_s$ is the mass source rate, and the conservation of charge arises naturally from Noether's theorem \citep{noether-235} applied to the gauge condition. 

The species contribution to the free momentum is $\mvec{K}_s \equiv \rho_s \mvec{V}_s$.  For $\mvec{P} \rightarrow 0$ the macroscopic Lorentz force reduces to \beq
\mvec{F}_\mrm{L} = \mvec{F}_\mrm{free} + \umu_0 (\mvec{M} \cdot \del) \mvec{H} - \umu_0 \epsi_0 (\psub{t} \mvec{M}) \times \mvec{E} \;,
\eeq and the species components $\mvec{F}_{s \mrm{L}}$ may be extracted.  With an influx $S_{s 1} \neq 0$ of mechanical momentum $\mvec{F}_{s 1} = \psub{t} \mvec{K}_{s 1}$, the balance of source, acceleration, and force for each species $\dot{\mvec{K}_s} = \mvec{F}_s$ is written \beq
\dot{\rho}_s \mvec{V}_s + \rho_s (\psub{t} + \mvec{V}_s \cdot \del) \mvec{V}_s = \mvec{F}_{s 1} + \mvec{F}_{s \mrm{G}} + \mvec{F}_{s \mrm{L}} + \mvec{F}_{s s} + \mvec{F}_{s k} \;,
\eeq where $\mvec{F}_{s \mrm{G}} = - \del \rho_s G$ accounts for gravitational acceleration in potential $G$, $\mvec{F}_{s \mrm{L}}$ is the generalized Lorentz force on the $s$ component of the fluid, the term $\mvec{F}_{s s} = - \divr (\mten{p}_s + \boldsymbol{\Pi}_s)$ represents intraspecies collisions, and the term $\mvec{F}_{s k} \equiv \sum_{s' \neq s} \mvec{F}_{s s'} \sim - \rho_s \nu_{s k} (\mvec{V}_s - \mvec{V}_k)$ represents interspecies collisions.  Using $\mvec{F}_{k s} = - \mvec{F}_{s k}$, their sum \beq
\sum_s \dot{\mvec{K}}_s = \mvec{F}_1 + \mvec{F}_\mrm{G} + \mvec{F}_\mrm{L} - \divr (\mten{p} + \boldsymbol{\Pi}) \;,
\eeq is the equation for the net conservation of momentum, which may also be written \beq
\dot{\mvec{K}}_\mrm{f} + \divr (m_e m_i / e^2 \rho_\mrm{m}) \mvec{J}_\mrm{f} \mvec{J}_\mrm{f} = \psub{t} (\mvec{K}_1 - \mvec{K}_{EH}) - \divr \left( \mten{p} + \boldsymbol{\Pi} + \mten{I} U_\mrm{G} - \mten{T}^{EB}_{EH} \right) \;,
\eeq where  $U_\mrm{G} = \rho_\mrm{m} G$ is the gravitational potential energy density, making explicit the appearance of the field momentum.  The term in $\mvec{J}_\mrm{f} \mvec{J}_\mrm{f}$ represents the free current's contribution to the convective force.

Noting that the hydrodynamic and thermodynamic pressures may differ by a factor of the bulk viscosity \citep{guyon2001}, our identification of (the divergence of) the pressure and viscosity tensors as a collision term relies on an argument presented by Woods \citep{woodsbook-04} that the mediator of the pressure force is the existence of scattering events and the observation that the Braginskii viscosity \citep{brag-1965} is written in terms of an energy density, a collision rate, and a shear tensor, $\boldsymbol{\Pi}_s = \sum_\alpha \eta^\alpha_s \mten{W}^\alpha_s$ for $\eta^\alpha_s \sim n_s T_s / \nu^\alpha_{ss}$ and $\mten{W}_s = \del \mvec{V}_s + (\del \mvec{V}_s)^\mrm{T} - (2/3) (\divr \mvec{V}_s) \mten{I}$.  We would like to express the interspecies collision term $\mvec{F}_{s k}$ along similar lines in order to account for the momentum transfer in detailed form, postponing such development for a later time.  The term for incoming force $\mvec{F}_1 = \sum_s \mvec{F}_{s 1} = \sum_s m_s \int \!\mvec{v}\, (\psub{t} f_s)_\mrm{S} \mrm{d}^3 \mvec{v}$ results from a source distribution $(\psub{t} f_s)_\mrm{S}$ which is not isotropic in the particle velocity $\mvec{v}_s = \mvec{V}_s + \mvec{u}_s$, thus yielding a net transfer of momentum.

The kinetic energy density for each species is written \beq
\mcal{E}_s = \Tr (\mten{p}_s + \boldsymbol{\Pi}_s + n_s m_s \mvec{V}_s \mvec{V}_s) / 2 = 3 p_s / 2 + n_s m_s V_s^2 / 2 \;,
\eeq where $p_s$ is the scalar pressure, whose evolution is described by the equation \beq
\dot{\mcal{E}_s} + \divr \mvec{q}_s = C_{s2} + S_{s2} + \mvec{V}_s \cdot \mvec{F}_s \;,
\eeq in terms of the heat flux density $\mvec{q}_s$, which may also be written \citep{fitzpatrick-notes} as \beq
\psub{t} \mcal{E}_s + \divr \mvec{Q}_s = C_{s2} + S_{s2} + \mvec{V}_s \cdot (\mvec{F}_s - \mvec{F}_{ss}) \;,
\eeq in terms of the energy flux density $\mvec{Q}_s = \mcal{E}_s \mvec{V}_s + (\mten{p}_s + \boldsymbol{\Pi}_s) \cdot \mvec{V}_s + \mvec{q}_s$.  Using $\sum_s (C_{s2} + \mvec{V}_s \cdot \mvec{F}_{sk}) = 0$ for elastic collisions, one may write the net energy conservation equation \bes
\sum_s \dot{\mcal{E}_s} + \divr \mvec{q} &=& \sum_s \left( C_{s2} + S_{s2} + \mvec{V}_s \cdot \dot{\mvec{K}_s} \right) \;, \\ 
 &=& \sum_s \left[ S_{s2} + \mvec{V}_s \cdot (\mvec{F}_s - \mvec{F}_{sk}) \right] \;,
\ees where the source moment $\sum_s S_{s2}$ gives the net influx of kinetic energy.  Closure is achieved by specifying the quantities $\mvec{q}_s$, $\mten{p}_s + \boldsymbol{\Pi}_s$, and $\nu_{ss'}$ in terms of the degrees of freedom $n_s$, $T_s$, and $\mvec{V}_s$ and the fields $\mvec{E}$ and $\mvec{H}$.

When evaluating the net kinetic energy of the plasma, one finds terms which depend on the current as well as the momentum.  Using the notation $\wt{\mvec{J}_\mrm{f}} \equiv \mvec{J}_\mrm{f} / e \rho_\mrm{m}$ and defining $
\mcal{E}_\mrm{f} \equiv 3 p / 2 + \rho_\mrm{m} ( V_\mrm{f}^2 + m_e m_i \wt{J_\mrm{f}}^2 ) / 2 \;,
$ one may write the rate of change of the net kinetic energy density as \bes
\sum_s \dot{\mcal{E}_s} = \dot{\mcal{E}_\mrm{f}} &+ &\divr 3 (m_e p_i - m_i p_e) \wt{\mvec{J}_\mrm{f}}/ 2 \\
 &+ &\divr \left\lbrace m_e m_i \rho_\mrm{m} \left[ \mvec{V}_\mrm{f} + (m_e - m_i) \wt{\mvec{J}_\mrm{f}}/ 2 \right] \cdot \wt{\mvec{J}_\mrm{f}} \right\rbrace \wt{\mvec{J}_\mrm{f}} \;,
\ees where $\dot{\mcal{E}_\mrm{f}} \equiv \psub{t} \mcal{E}_\mrm{f} + \divr \mcal{E}_\mrm{f} \mvec{V}_\mrm{f}$.  Net energy conservation is best interpreted as the dynamical equation for the pressure, as it explicitly relates $\psub{t} p$ to other quantities determined elsewhere.

\section{Ohm's law}\label{sec:ohmslaw}
One final piece of the puzzle remains: determining the relationship between the electric field and the current, which in its simplest form reduces to the linear relationship $\mvec{J}_\mrm{f} = \sigma \mvec{E}$.  For a medium in which all species of charge carriers may flow, the relationship is not so simple.  The theory provides for two types of coupling, mass and charge, by which forces may act on particles, and while momentum is driven (along $\mvec{H}$) by forces acting on ions and electrons in the same direction, current is created by forces which drive them apart.  (Perpendicular to $\mvec{H}$, life gets a bit more complicated.)  Consequently, the generalized Ohm's law is given by (the sum of) the ion momentum conservation equation(s) minus the electron equation, which may also be thought of as an equation for positrons going backwards in time \citep{mattuck-1976}.

Many authors derive Ohm's law  ``by taking a particular linear combination of the fluid equations'' \citep{dendybook-93} where the ion equation is reduced by a factor of $m_e / m_i$, retaining only terms which survive the limit $m_e \ll m_i$.  We argue that such procedure does not tell the whole story, effectively neglecting the remainder of the ion contribution $(1-m_e/m_i)$, nor does it respect the unwritten factor of units inherent in physical equations (as opposed to purely mathematical ones).  Before adding or subtracting two physical quantities, they must be expressed in the same units, including prefix.  The extraneous mass factors applied when taking $m_e \mvec{F}_i - m_i \mvec{F}_e$ may just as well be replaced by ``milli'' and ``mega'' as prefix on the originally common units, so that the difference has numerically subtracted mega-Newtons from milli-Newtons.  Just as the free momentum equation relies on the net force balance $\dot{\mvec{K}}_i + \dot{\mvec{K}}_e = \mvec{F}_i + \mvec{F}_e$, so does the free current equation rely on the net force difference $\dot{\mvec{K}}_i - \dot{\mvec{K}}_e = \mvec{F}_i - \mvec{F}_e$, which through substitution yields the equation for $\dot{\mvec{J}_\mrm{f}}$.  From previous derivation \citep{rwj-pop01}, we note that the extraneous factors institute a simplification of the equation through the artificial cancellation of certain terms, particularly in the convective derivative, thus revealing their likely motivation.

\section{Discriminatory apparatus}\label{sec:eval}
\begin{figure}
\includegraphics[]{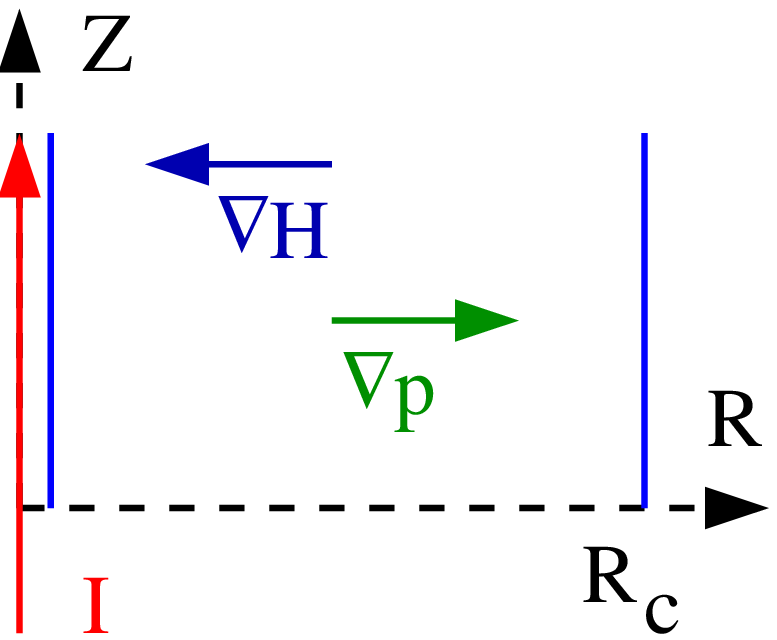}
\caption{Plasma chamber surrounding a current carrying wire. A pressure gradient should develop to oppose the magnetic polarization force.}
\label{fig:A}
\end{figure}

After so much development, we restrict ourselves to a single evaluation.  Consider an annular plasma chamber in $(R,\phi,Z)$ coordinates surrounding a conductive wire along the $Z$-axis which is insulated from the plasma, as in Fig.~\ref{fig:A}.  Supposing the plasma to be stationary and carrying no free current, the net force balance equation reduces to \bes
\del p = \umu_0 (\mvec{M} \cdot \del) \mvec{H} &= &- \umu_0 (M/H) (\mvec{H} \cdot \del) \mvec{H} \;, \\
 &= &- \umu_0 (M/H) \del H^2 / 2 \;,
\ees in the absence of gravitational acceleration $\mvec{g}$, where we have used $\psub{t} = 0$ and $\curl \mvec{H} = 0$ in the plasma region.  Without the media contribution to the Lorentz force, one has simply $\del p = 0$ thus a constant pressure in the chamber, and the current in the wire should have no effect on the plasma, thus offering a chance for an experimental determination of its existence.  A thorough analysis would need to account for the momentum transfer by the electric field during the period when the current is changing.

Sufficiently far from the ends of the wire, its field is given by Ampere's law as $\mvec{H}^\infty_\mrm{wire} = \phihat H_0 / R$ for $H_0 = I_Z / 2 \upi$, where $I_Z$ is the total current, and its magnitude has the gradient $\del H^2 / 2 = - \Rhat H_0^2 / R^3$.  Using $\ptild \equiv p / \umu_0$, the net force balance becomes the scalar equation \bes
\psub{R} \ptild &= &- \oover{4} \left( 1 - \sqrt{1 - 4 \ptild / H_\phi^2} \right) \psub{R} H_\phi^2 \;, \\
 &= &\half \left( 1 - \sqrt{1 - 4 \ptild R^2 / H_0^2 } \right) H_0^2 R^{-3} \;,
\ees which one may solve for the pressure profile given values for the current $I_Z$ and the pressure at the outer chamber wall $p_c$.

\begin{figure}
\includegraphics[]{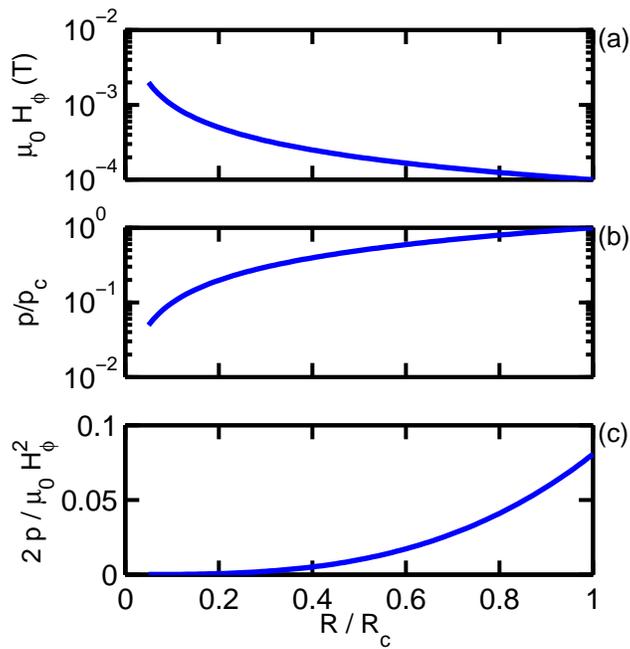}
\caption{Solution for a plasma chamber with outer radius of $R_c=100$~mm and inner radius of 5~mm surrounding a wire at 0~mm carrying 50~A current upward. (a) Magnetic field profile. (b) Normalized pressure profile for $p_c = 0.32$~mPa. (c) The $\beta_H$ profile remains below the limit of 0.5 at the edge of the chamber.}
\label{fig:B}
\end{figure}

For a plasma chamber with $R_c = 100$~mm outer radius and inner radius of 5~mm surrounding a current of 50~A along $\Zhat$ at $R=0$, we calculate the $\mvec{H}$ field profile in Fig.~\ref{fig:B}(a).  Assuming a common temperature of 100~eV and a net edge density of $2\times 10^{13}$~$\mrm{m}^{-3}$, the pressure profile in Fig.~\ref{fig:B}(b) is normalized by the pressure at the chamber wall $p_c = 0.32$~mPa.  In the presence of $\mvec{F}_\mrm{media}$, the magnetic polarization force is balanced by a pressure gradient such that the pressure at the centre is reduced by a factor $\sim$50 relative to its value at the edge.  Without that force, there should be no pressure gradient.  The location of maximum $\beta_H$ is at the outer wall $R_c$, and for these parameters it remains below its limit of 0.5 as shown in Fig.~\ref{fig:B}(c).  The upshot is that experimentalists should be able to construct an apparatus which can measure the effect predicted by the theory, requiring verification that fluid, diamagnetic material should be repelled from a stronger field region.

\section{Conclusions and outlook}\label{sec:conc}
In summary, we have found that the Lorentz force on the free source densities may be incorporated self-consistently with the media force proposed by Mansuripur \citep{mansuripur-1608,mansuripur-1619} and investigated by Louden {\it et al} \citep{loudon-063802,louden-1411855L} to yield the macroscopic electromagnetic stress tensor in terms of the fields $\mvec{D}$, $\mvec{B}$, $\mvec{E}$, and $\mvec{H}$ without requirements on the detailed form of the constitutive relations.  The nonlinear model for plasma magnetization is given a solution which displays a limit on the ratio of kinetic to magnetic pressure.  The macroscopic conservation equations for a fully ionized medium are expressed in terms of the free momentum and current, making explicit the contribution of each to the kinetic energy density.

With restriction to a stationary plasma without free current and neglecting gravity, conservation of momentum requires a pressure gradient develop to oppose the magnetic polarization force.  An evaluation for parameters which may be accessible to experimentalists suggests a measurable difference between the central and edge pressures for a chamber surrounding a steady current, indicating the presence of the media contribution to the generalized Lorentz force.

The complete form of the macroscopic Lorentz force bears directly on the physics of plasmas.  Treatments by the fluid description need to respect the dielectric and diamagnetic properties of the medium as encoded in the constitutive relations.  The model presented here is not the only one available, and there is something to be said \citep{rwj-cpp03} for using the expression $[\curl (\mvec{H} + \mvec{M})] \times (\mvec{H} + \mvec{M}) + \del (\mvec{M} \cdot \mvec{H})$ as the magnetic contribution to the Lorentz force.  Variations in a theoretical development are no bad thing, as they allow for the possibility to discriminate between competing models.  We consider the full effect of plasma magnetization to be an open question, requiring a conspiracy between theorists and experimentalists for its determination.





\newpage

\end{document}